\documentclass[twocolumn,amssymb]{revtex4}
\usepackage{graphicx}
\usepackage{bm}
\begin{document}

\title{Large scale inhomogeneity of inertial particles in turbulent flow}


\author{G. Boffetta}
\affiliation{Dipartimento di Fisica Generale and INFM, Universit\`a di Torino,
via P. Giuria 1, 10125 Torino, Italy, \\
and CNR-ISAC - Sezione di Torino, corso Fiume 4, 10133 Torino, Italy}
\author{F. De Lillo}
\affiliation{Dipartimento di Fisica Generale and INFM, Universit\`a di Torino,
via P. Giuria 1, 10125 Torino, Italy, \\
and CNR-ISAC - Sezione di Torino, corso Fiume 4, 10133 Torino, Italy}
\author{A. Gamba}
\affiliation{Dipartimento di Matematica, Politecnico di Torino,
corso Duca degli Abruzzi 24, 10129 Torino, Italy}
\date{\today}

\begin{abstract}
Preferential concentration of inertial particles in 
turbulent flow is studied by high resolution direct numerical simulations
of two-dimensional turbulence. 
The formation of network-like regions of high particle density, 
characterized by a length scale which depends on the Stokes number 
of inertial particles, is observed.
At smaller scales, the size of empty regions appears to be distributed 
according to a scaling law. 
\end{abstract}

\maketitle

The transport of inertial particles in fluids displays properties
typical of compressible motion even in incompressible flows.
This is a consequence of the difference of density between 
particles and fluid.
The most peculiar effect is the spontaneous generation of 
inhomogeneity out of an initially homogeneous distribution.
The clustering of inertial particles has important
physical applications, from rain generation \cite{FFS02}, to 
pollutant distribution and combustion~\cite{DWM96}.
Starting from the first examples in laminar flow \cite{Maxey87},
it is now demonstrated both numerically \cite{SE91,WM93,HC01}
and experimentally \cite{FKE94,ACHL02} that also in fully developed
turbulence there is a tendency of inertial particles to 
form small scale clusters.
The parameter characterizing the effect of inertia is 
the Stokes number $\mathrm{St}$, defined as the ratio between the
particle viscous response time $\tau_\mathrm{s}$ and a characteristic time
of the flow $\tau_\mathrm{v}$. 
In the limit $\mathrm{St} \to 0$ inertial
particles recover the motion of fluid particles and no
clusterization is expected. In the opposite limit $\mathrm{St} \to \infty$
particles become less and less influenced by the velocity field.
The most interesting situation is observed for intermediate 
values of $\mathrm{St}$ where strong clusterization is observed 
\cite{FKE94,HC01}.

In the case of a smooth velocity field the Eulerian characteristic time
$\tau_\mathrm{v}$ is a well defined quantity as it can be identified
with the inverse Lyapunov exponent $\tau_\mathrm{v}=\lambda_{1}^{-1}$
of fluid trajectories.
In this case some general theoretical predictions are possible
\cite{EKR96,BFF01} such as the exponential growth of high 
order concentration moments. Detailed numerical simulations
in a chaotic random flow have shown maximal clusterization
(measured in terms of the dimension of the Lagrangian attractor)
for a value $\mathrm{St}\simeq 0.1$ \cite{Bec03}.

In the case of turbulent flow, where the velocity field is not smooth,
a simple scaling argument suggests that maximal compressibility
effects are produced by the smallest, dissipative scales \cite{BFF01}.
Nevertheless, for sufficiently large values of $\mathrm{St}$, 
the particle response time introduces
a characteristic scale in the inertial range which breaks the
scale invariance of the velocity field and produces, as we will see,
large scales inhomogeneity in particle distribution.

\begin{figure*}[htb]
\centerline{\hspace{-0.0cm}
\includegraphics[draft=false, scale=0.75]{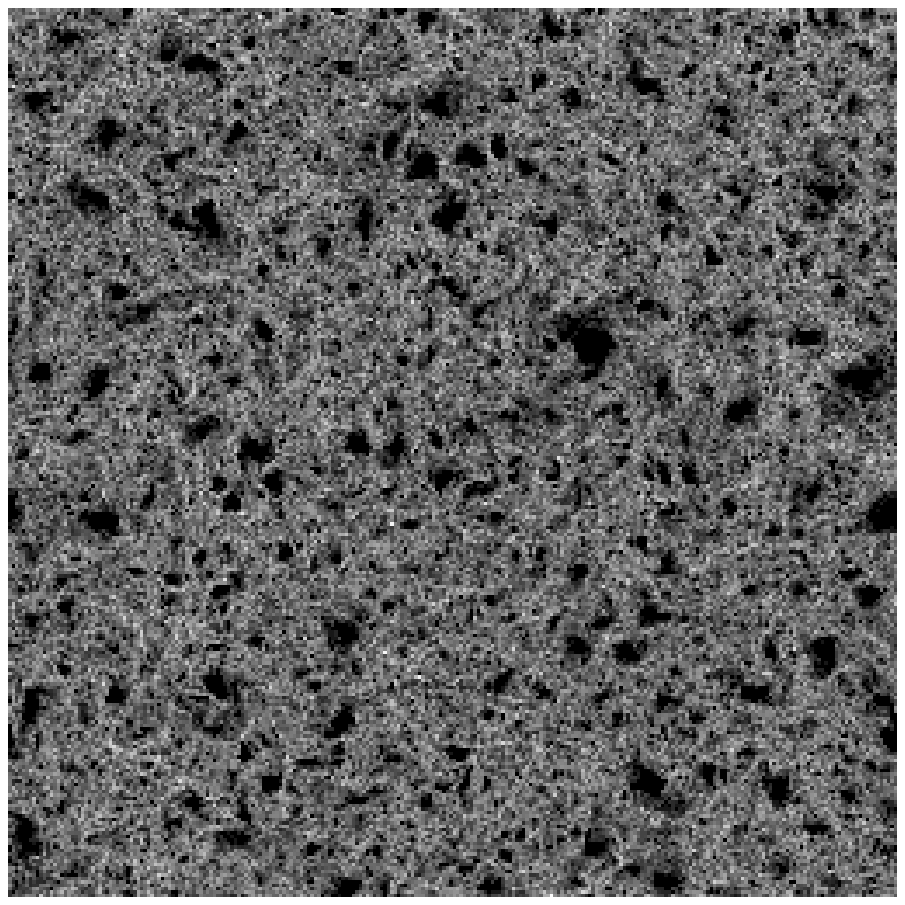}
\includegraphics[draft=false, scale=0.75]{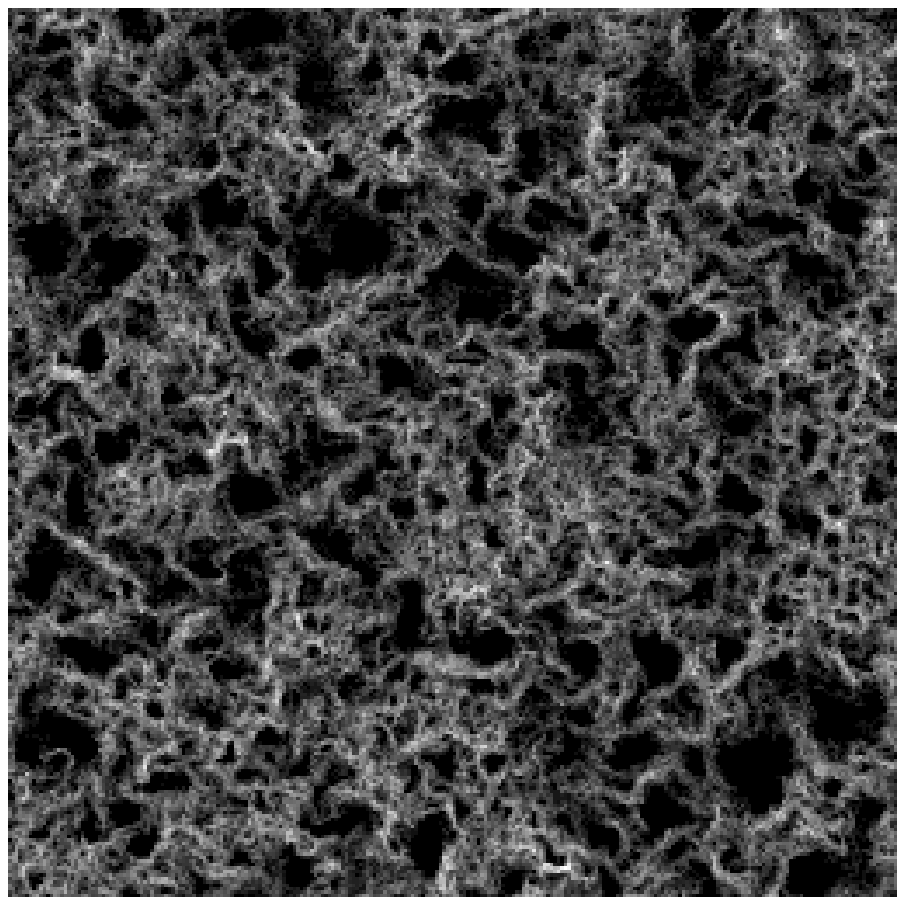}}
\caption{
Snapshots of particle concentrations taken at the same time
in stationary condition for two realizations with 
different Stokes numbers \(\mathrm{St}=0.12\) (left) and 
\(\mathrm{St}=1.2\) (right) started from identical initial conditions
and advected by the same turbulent flow. 
The number of particles in both cases is $1024^2$.
Gray levels correspond to particle concentration.
Turbulent velocity is obtained by integration of two-dimensional
Navier--Stokes equation by a standard pseudo-spectral code on 
doubly periodic unit square domain at resolution $N^2=1024^2$.
}
\label{fig1}
\end{figure*}

The motion of a spherical particle in an incompressible flow,
when the size of the particle is so small
that the surrounding flow can be approximated by a Stokes flow,
is governed by the set of equations~\cite{MR87}
\begin{equation}
\left\{
\begin{array}{l}
\dot{\bm x} = {\bm v} \\
\\
\dot{\bm v} = 
- {\displaystyle 1 \over \displaystyle\tau_\mathrm{s}} \left[{\bm v} 
- {\bm u}({\bm x(t)},t) \right]
+\beta{\displaystyle\mathrm{d}\over\displaystyle\mathrm{d}t}{\bm u}
({\bm x(t)},t)
\end{array}
\right.
\label{eq:1}
\end{equation}
where ${\bm v}$ represents the Lagrangian velocity of the 
particle, \(\beta=3\rho_0/(\rho_0+2\rho)\) and  $\rho$ and $\rho_0$
are the density of particle and fluid respectively.
In (\ref{eq:1}) ${\bm u}({\bm x},t)$ represents
the velocity field whose evolution is
given by Navier-Stokes equations
\begin{equation}
{\partial {\bm u} \over \partial t} + {\bm u} \cdot {\bm \nabla} {\bm u} =
-  {\bm \nabla} p + \nu \Delta {\bm u} + {\bm f}
\label{eq:2}
\end{equation}
In what follows, we will consider the limit heavy particles such that
\(\beta\simeq 0\). In this limit
it is easy to show that the Lagrangian velocity possesses
a compressible part \cite{BFF01}: expanding (\ref{eq:1}) to 
first order in $\tau_\mathrm{s}$ and using ${\bm \nabla} \cdot {\bm u}=0$,
one obtains, from (\ref{eq:2})
\begin{equation}
{\bm \nabla} \cdot {\bm v} \simeq - \tau_\mathrm{s} {\bm \nabla} \cdot \left( 
{\bm u} \cdot {\bm \nabla} {\bm u} \right) \ne 0
\label{eq:3}
\end{equation}
From (\ref{eq:3}) it is possible to give a dimensional estimation of 
the relative importance of the compressible
part for a turbulent velocity field with scaling exponent $h$,
$\delta_{\ell} u \sim U (\ell/L)^h$, $L$ and $U$ being a characteristic
large scale and velocity.
The scaling exponent for the compressible component of ${\bm v}$ is
$\delta_{\ell} v \sim U(\ell/L)^{2h-1}$ and thus the relative compressibility
scales as $(\ell/L)^{h-1}$, i.e. reaches the maximum value at the 
viscous scale \cite{BFF01}.
Nevertheless, we will see that the presence of inertial range of 
scales in the turbulent flow generates
large scale structures in the particle distribution at large St.

In this letter we address the problem of transport of heavy inertial
particles in fully developed two-dimensional turbulence in the inverse
cascade regime. 
High resolution direct numerical simulations of 
two-dimensional Navier-Stokes equation with white in time, random 
forcing ${\bm f}$ localized at small scales $\ell_\mathrm{f}$ have been
performed.
As customary, a friction term $-\alpha {\bm u}$ is added to (\ref{eq:2})
in order to extract energy from the system at the friction scale 
$\ell_\mathrm{fr} \sim \varepsilon^{1/2} \alpha^{-3/2}$. The intermediate 
scales $\ell_\mathrm{f} \ll \ell \ll \ell_\mathrm{fr}$ define the inertial
range in which Kolmogorov scaling $\delta_{\ell} u \sim U (\ell/L)^{1/3}$
is observed \cite{BCV00}.
Lagrangian tracers are placed at random with initial zero velocity and
integrated according to (\ref{eq:1}) with a given $\tau_\mathrm{s}$.
After a scratch run long several $\tau_\mathrm{s}$, Lagrangian statistics 
is accumulated for typically some tens of $\tau_\mathrm{s}$.
Stokes time is made dimensionless by rescaling with the Lagrangian
Lyapunov exponent of fluid particles, 
$\mathrm{St} \equiv \lambda_{1} \tau_\mathrm{s}$.
Fig.~\ref{fig1} shows typical distributions of inertial tracers
in stationary conditions at different values of $\mathrm{St}$,
obtained starting from the same initial random distribution.
One observes in both cases strong inhomogeneous distributions with 
empty ``holes'', in the second case on much larger scales.

\begin{figure}[ht]
\centerline{\hspace{-0.0cm}
\includegraphics[draft=false, scale=0.68]{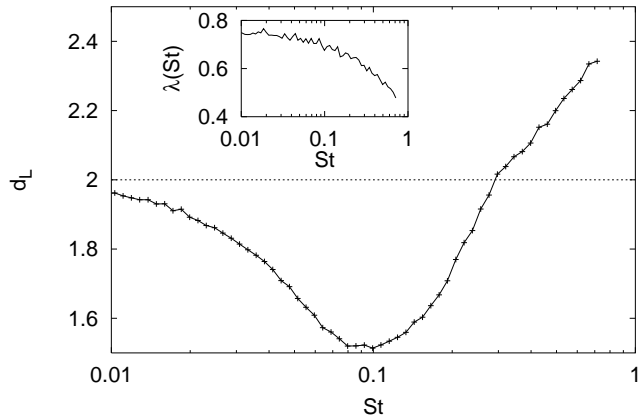}}
\caption{Lyapunov dimension for heavy particles in two-dimensional 
turbulence as a function of the Stokes number. 
Inset: the first Lyapunov exponent $\lambda_1$.
}
\label{fig2}
\end{figure}

Maximum compressibility effects are expected at
small scales and can be described by the 
Lyapunov spectrum for inertial particles. We recall that 
for a generic dynamical system
the sum of the Lyapunov exponents
gives the exponential rate of expansion (or contraction) of the hypervolume
in phase space. In our case, from (\ref{eq:1}) we have 
$\sum_{i=1,4} \lambda_i = -2/\tau_\mathrm{s}$, thus volumes are contracted
at a constant rate. Let us observe that when $\tau_\mathrm{s}\to \infty$,
phase space contraction rate vanishes, and we thus expect 
less clusterization.
As a consequence of the structure of (\ref{eq:1}) we find that two 
Lyapunov exponents are close to $-1/\tau_\mathrm{s}$, representing the 
rate of adjustment of Lagrangian velocity to the Eulerian one. 
The first Lyapunov exponent is found positive,
as the trajectories are chaotic and the second, negative,
determines the dimension of the attractor according to the definition
of Lyapunov dimension \cite{ott}
\begin{equation}
d_\mathrm{L} = K + {\sum_{i=1}^K \lambda_i \over |\lambda_{K+1}|}
\label{eq:4}
\end{equation}
where $K$ is defined as the largest integer such that 
$\sum_{i=1}^K \lambda_i \ge 0$. 
In the inset of Fig.~\ref{fig2} we show the dependence of the first
Lyapunov exponent on Stokes number. 
As $\mathrm{St}$ increases, $\lambda_{1}(\mathrm{St})$ 
decreases monotonically from the neutral value $\lambda_{1}(0) \simeq 0.72$.
The behavior of Lyapunov dimension is also shown in Fig.~\ref{fig2}.
In the limit $\mathrm{St} \to 0$, particles become neutral and thus one
recovers the homogeneous distribution with $d_\mathrm{L}=2$. The presence
of a minimum around $\mathrm{St} \simeq 0.1$ was already discussed in the case 
of smooth flow \cite{Bec03} and indicates a value for which 
compressibility effects are maximum. 
We remark that the curve $d_{L}(St)$ of Fig.~\ref{fig2} is
almost identical to the one obtained in synthetic smooth flow
\cite{Bec03}, and is thus insensitive to the presence of
the hierarchy of scales typical of a turbulent flow.

In the turbulent scenario, 
instead of becoming more homogeneous, for increasing \(\mathrm{St}\)
the inertial particle distribution develops structures on larger scales,
as is evident in Fig.~\ref{fig1}.
The most evident dishomogeneities are related to the presence of empty
regions (``holes'') on different scales. We have thus studied the hole 
statistics at varying Stokes number.

We have performed a coarse graining of the system by dividing it into
small boxes forming the sites of a square lattice and counting the
number of particles contained in each small box.
From this coarse grained density we have computed the probability density
function of holes, defined as connected regions of empty boxes.

\begin{figure}[ht]
\centerline{\hspace{-0.0cm}
\includegraphics[draft=false, scale=0.7]{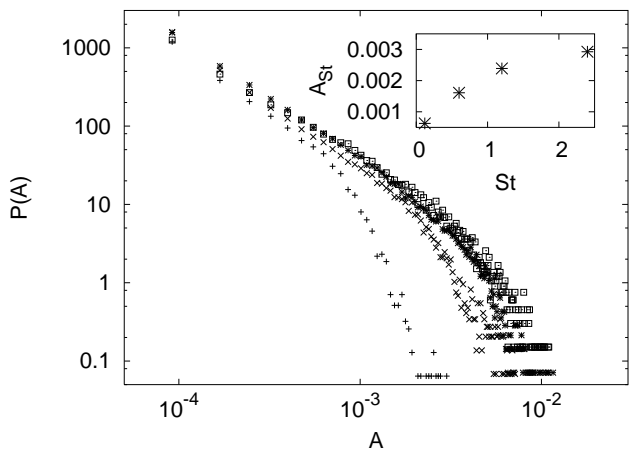}}
\caption{Probability density functions (pdf) of hole areas,
normalized with the area of the box, for 
$\mathrm{St}=0.12$ ($+$), $\mathrm{St}=0.6$ ($\times$), 
$\mathrm{St}=1.2$ ($*$) and $\mathrm{St}=2.4$ ($\square$).
Holes are defined as connected regions of coarse grained distribution
with zero density. Probability density functions are
computed over $100$ independent realizations of $1024^2$ particles
each. In the inset we show the dependence of the cutoff area
$A_\mathrm{St}$ (defined by the condition that $1\%$ of holes has larger
area) on $\mathrm{St}$. The robustness of the hole area census
with respect to particle statistics have been checked by increasing 
the number of tracers up to $4~\times~10^{6}$.
}
\label{fig3}
\end{figure}

Probability density functions (pdf) of hole areas are shown in 
Fig.~\ref{fig3} for different Stokes numbers. 
The hole distributions follow a power law 
with an exponent \(-1.8\pm0.2\) up to an exponential 
cutoff at a scale \(A_\mathrm{St}\) which
moves to larger sizes with $\mathrm{St}$, as shown in the inset
of Fig.~\ref{fig3}.
At variance with the smooth flow case, in which particle density
recovers homogeneity for $\mathrm{St}>0.1$ \cite{Bec03},
in the turbulent case inhomogeneities are pushed to larger scales 
when increasing $\mathrm{St}$.

The hole area pdf is independent on the number of particles
used in the simulations.
This is a non trivial property, reflecting the
fact that inertial particles cluster on network-like structures
where a clear-cut distinction between empty regions and particle-rich regions 
can be observed.
We have also verified that the choice of the small coarse graining scale does
not modify the hole area pdf at larger scales. 

\begin{figure}[ht]
\centerline{\hspace{-0.0cm}
\includegraphics[draft=false, scale=0.7]{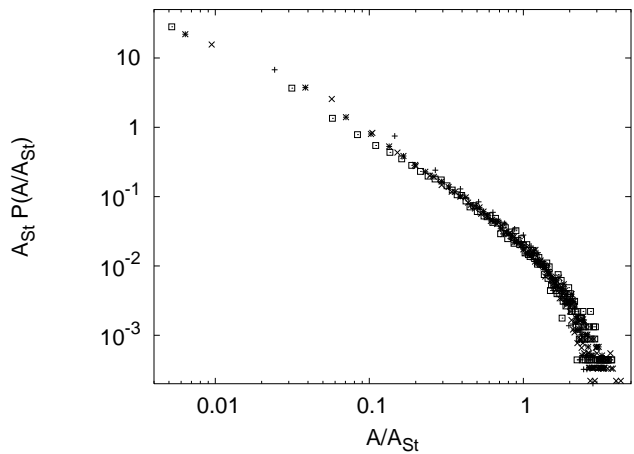}}
\caption{Probability density functions of hole areas rescaled
with the cutoff areas $A_\mathrm{St}$ computed from Fig.\ref{fig3}.
Symbols as in Fig.\ref{fig3}.
}
\label{fig4}
\end{figure}

These results support the following physical picture.
According to (\ref{eq:1}), heavy particles filter out velocity
fluctuations on timescales shorter than $\tau_{S}$. We may thus
expect that inhomogeneity does appear at a scale $\ell_{S}$ whose
characteristic time is of order of $\tau_{S}$.
For a velocity field with a roughness exponent $h \le 1$, this 
would lead to 
$A_\mathrm{St} \sim \mathrm{St}^{2/(1-h)}$, i.e. 
to the increase of hole areas 
with $\mathrm{St}$, in qualitative agreement with Fig.~\ref{fig3}.
Clusterization of inertial particles in turbulence appears to be
a selfsimilar process up the cutoff scale $A_\mathrm{St}$. 
Indeed the pdfs of hole areas collapse when areas
are rescaled with the cutoff area, as shown in Fig.~\ref{fig4}.

\begin{figure}[ht]
\centerline{\hspace{-0.0cm}
\includegraphics[draft=false, scale=0.7]{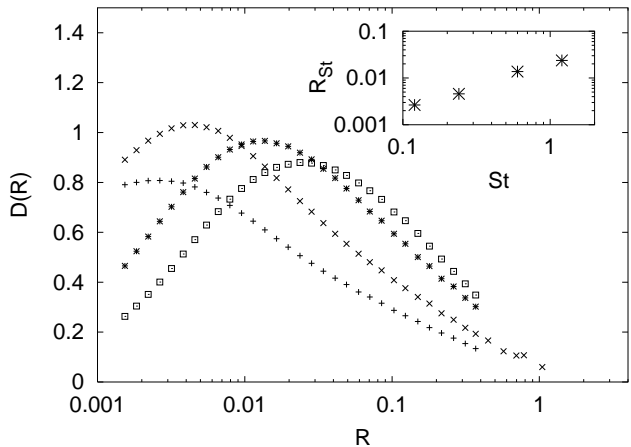}}
\caption{Deviation of the inertial particle distribution from 
Poisson law as a function of the coarse grain scale for different
$St$. The maximum of $D(R)$ defines the clusterization scale
$R_{St}$, whose dependence on $St$ is shown in the inset.
}
\label{fig5}
\end{figure}

A natural measure of the degree of clusterization of inertial particles
which has been considered in the literature~\cite{SE91,HC01,FKE94} is the
deviation of the particle density distribution from Poisson law,
$D(R)=(\sigma_{R}-\sigma_{R} ^{(P)})/\lambda_{R}$,
where $\lambda_{R}$ and $\sigma_{R}$ represent the mean and standard
deviation of particle distribution coarse grained at scale $R$ and
$\sigma_{R}^{(P)}$ is the corresponding standard deviation for a Poisson
distribution.
The maximum of the deviation $D(R)$, which depends on $St$
(see Fig.~\ref{fig5}) is then used for defining the typical scale of 
clusters, $R_{St}$.
We have found that this definition of $D(R)$ is strongly sensitive 
to particle statistics, nevertheless, the 
increase of $R_{St}$ on $St$ (inset of Fig.~\ref{fig5}) seems 
to be a robust result.

The presence of structures in the inertial particle distribution is often
attributed~\cite{SE91} to the fact that heavy particles are expelled from 
vortical regions.
Although structures are related to the presence
of many active scales in the turbulent flow, one should recall 
that in 2D turbulence, as a consequence of the direct vorticity
cascade, vorticity is concentrated at the small scales~\cite{BCV00}
and no large scale coherent structures appear.
Holes emerge as a result of the delayed dynamics~(\ref{eq:1}), 
which filters the scales of the underlying
turbulent flow characterized by times of the order of the 
Stokes time $\tau_{\mathrm{s}}$.

As discussed before, clustering occurs also in synthetic flows
where a hierarchy of time scales is absent, just as a consequence
of the dissipative character of the motion \cite{Bec03}.
However, it appears from our simulations that to fully understand
the geometry of inertial particle distribution in a turbulent flow
the presence of structures characterized by a large set of time scales
cannot be ignored.

We conclude that the geometry of inertial particle clusters in developed
turbulence is controlled both by the dissipative effective dynamics of
the particle motion at small scales, and by the tendency of inertial
particles to filter the active scales characterized by times
of the order of the characteristic relaxation time of the particles.
A full understanding of the geometry of particle clusters in developed 
turbulence is particularly relevant for several applications, such
as coalescence processes or chemical reactions.
The reaction rate of two chemical species is a function of
their concentration.
When particles of different species are transported by the same 
turbulent flow, their local concentrations are not independent
and the presence of large scale correlations in the particle 
distribution can in principle influence the reaction velocity.

This work was supported by MIUR-Cofin2001 contract 2001023848.
We acknowledge the allocation of computer resources from INFM
``Progetto Calcolo Parallelo''.

%


\begin{thebibliography}{99}


\bibitem{FFS02}
G. Falkovich and A. Fouxon and M.G. Stepanov,
Nature {\bf 419}, 151 (2002).

\bibitem{DWM96}
R.W. Dibble, J. Warnatz and U. Maas,
Combustion: physical and chemical fundamentals, modelling and simulations,
experiments, pollutant formation.
Springer, New York, 1996

\bibitem{Maxey87}
M. Maxey, J. Fluid Mech. {\bf 174}, 441 (1987).

\bibitem{SE91}
K.D. Squires and J.K Eaton, Phys. Fluids A {\bf 3}, 1169 (1991).

\bibitem{WM93}
L.P. Wang and M. Maxey, J. Fluid Mech. {\bf 256}, 27 (1993).

\bibitem{HC01}
R.C. Hogan and J.N Cuzzi, Phys. Fluids {\bf 13}, 2938 (2001).

\bibitem{FKE94}
J.R. Fessler, J.D. Kulick and J.K. Eaton, Phys. Fluids {\bf 6}, 3742 (1994).

\bibitem{ACHL02}
A. Aliseda, A. Cartellier, F. Hainaux and J.C. Lasheras,
J. Fluid Mech. {\bf 468}, 77 (2002).

\bibitem{EKR96}
T. Elperin, N. Kleeorin and I. Rogachevskii, 
Phys. Rev. Lett. {\bf 77}, 5373 (1996).

\bibitem{BFF01}
E. Balkovsky, G. Falkovich and A. Fouxon,
Phys. Rev. Lett. {\bf 86}, 2790 (2001).

\bibitem{Bec03}
M.J. Bec, Fractal clustering of inertial particles in random flow,
Phys. Fluids {\bf 16}, L81 (2003).

\bibitem{MR87}
M. Maxey and J. Riley, Phys. Fluids {\bf 26}, 883 (1983).

\bibitem{BCV00}
G. Boffetta, A. Celani and M. Vergassola, Phys. Rev. E {\bf 61}, R29 (2000).

\bibitem{ott}
E. Ott, {\it Chaos in Dynamical Systems}, Cambridge University Press (1993).



%
%
%
%
%
\end{thebibliography}
\end{document}